\documentclass{article}
\usepackage{graphicx} % Required for inserting images
\usepackage{latexsym}
\usepackage{amscd}
\usepackage[utf8]{inputenc}
\usepackage[T1, T2A]{fontenc}
\usepackage[english]{babel}
\usepackage{amsmath}
\usepackage{amssymb}
\usepackage{verbatim}
\usepackage{xcolor}
\usepackage{indentfirst}
\usepackage{import}
\usepackage{xifthen}
\usepackage{pdfpages}
\usepackage{transparent}
\usepackage{appendix}
\usepackage{subcaption}
\usepackage{dirtytalk}

\newcommand{\incfig}[1]{%
    \def\svgwidth{\columnwidth}
\begingroup%
  \makeatletter%
  \providecommand\color[2][]{%
    \errmessage{(Inkscape) Color is used for the text in Inkscape, but the package 'color.sty' is not loaded}%
    \renewcommand\color[2][]{}%
  }%
  \providecommand\transparent[1]{%
    \errmessage{(Inkscape) Transparency is used (non-zero) for the text in Inkscape, but the package 'transparent.sty' is not loaded}%
    \renewcommand\transparent[1]{}%
  }%
  \providecommand\rotatebox[2]{#2}%
  \newcommand*\fsize{\dimexpr\f@size pt\relax}%
  \newcommand*\lineheight[1]{\fontsize{\fsize}{#1\fsize}\selectfont}%
  \ifx\svgwidth\undefined%
    \setlength{\unitlength}{197.26685291bp}%
    \ifx\svgscale\undefined%
      \relax%
    \else%
      \setlength{\unitlength}{\unitlength * \real{\svgscale}}%
    \fi%
  \else%
    \setlength{\unitlength}{\svgwidth}%
  \fi%
  \global\let\svgwidth\undefined%
  \global\let\svgscale\undefined%
  \makeatother%
  \begin{picture}(1,0.89133034)%
    \lineheight{1}%
    \setlength\tabcolsep{0pt}%
    \put(0,0){\includegraphics[width=\unitlength,page=1]{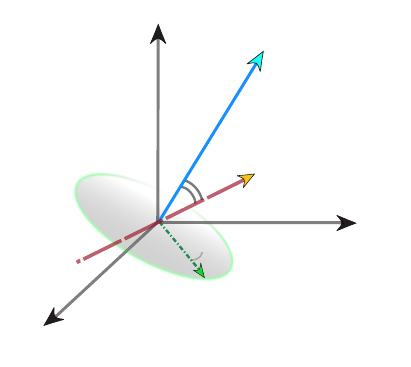}}%
    \put(0.63555055,0.46141542){\color[rgb]{0,0,0}\transparent{0.84815902}\makebox(0,0)[lt]{\lineheight{1.25}\smash{\begin{tabular}[t]{l}$\textbf{\Huge{u}}$\end{tabular}}}}%
    \put(0.65301327,0.76058044){\color[rgb]{0,0,0}\transparent{0.84815902}\makebox(0,0)[lt]{\lineheight{1.25}\smash{\begin{tabular}[t]{l}$\textbf{\Huge{v}}$\end{tabular}}}}%
    \put(0.12036283,0.06179387){\color[rgb]{0,0,0}\transparent{0.84815902}\makebox(0,0)[lt]{\lineheight{1.25}\smash{\begin{tabular}[t]{l}\Huge{$\mathbf{e}_{\alpha}$}\end{tabular}}}}%
    \put(0.48320426,0.15828874){\color[rgb]{0,0,0}\transparent{0.84815902}\makebox(0,0)[lt]{\lineheight{1.25}\smash{\begin{tabular}[t]{l}\Huge{$\mathbf{k}_{\perp}$}\end{tabular}}}}%
    \put(0.37094195,0.29283804){\color[rgb]{0,0,0}\transparent{0.84815902}\makebox(0,0)[lt]{\lineheight{1.25}\smash{\begin{tabular}[t]{l}\Large{\textbf{O}}\end{tabular}}}}%
    \put(0,0){\includegraphics[width=\unitlength,page=2]{Coordinat_system_mod.pdf}}%
    \put(0.50620978,0.457507){\color[rgb]{0,0,0}\transparent{0.93108702}\makebox(0,0)[t]{\lineheight{1.25}\smash{\begin{tabular}[t]{c}\Huge{$\theta$}\end{tabular}}}}%
    \put(0.50244866,0.24071694){\color[rgb]{0,0,0}\transparent{0.93108702}\makebox(0,0)[lt]{\lineheight{1.25}\smash{\begin{tabular}[t]{l}\Large{$\phi$}\end{tabular}}}}%
    \put(0.4023943,0.81952419){\color[rgb]{0,0,0}\transparent{0.84815902}\makebox(0,0)[lt]{\lineheight{1.25}\smash{\begin{tabular}[t]{l}\Huge{$\mathbf{e}_{\gamma}$}\end{tabular}}}}%
    \put(0.85276768,0.30799426){\color[rgb]{0,0,0}\transparent{0.84815902}\makebox(0,0)[lt]{\lineheight{1.25}\smash{\begin{tabular}[t]{l}\Huge{$\mathbf{e}_{\beta}$}\end{tabular}}}}%
    \put(0,0){\includegraphics[width=\unitlength,page=3]{Coordinat_system_mod.pdf}}%
  \end{picture}%
\endgroup%

}

\newcommand{\incfg}[1]{%
    \def\svgwidth{\columnwidth}
\begingroup%
  \makeatletter%
  \providecommand\color[2][]{%
    \errmessage{(Inkscape) Color is used for the text in Inkscape, but the package 'color.sty' is not loaded}%
    \renewcommand\color[2][]{}%
  }%
  \providecommand\transparent[1]{%
    \errmessage{(Inkscape) Transparency is used (non-zero) for the text in Inkscape, but the package 'transparent.sty' is not loaded}%
    \renewcommand\transparent[1]{}%
  }%
  \providecommand\rotatebox[2]{#2}%
  \newcommand*\fsize{\dimexpr\f@size pt\relax}%
  \newcommand*\lineheight[1]{\fontsize{\fsize}{#1\fsize}\selectfont}%
  \ifx\svgwidth\undefined%
    \setlength{\unitlength}{665.84358756bp}%
    \ifx\svgscale\undefined%
      \relax%
    \else%
      \setlength{\unitlength}{\unitlength * \real{\svgscale}}%
    \fi%
  \else%
    \setlength{\unitlength}{\svgwidth}%
  \fi%
  \global\let\svgwidth\undefined%
  \global\let\svgscale\undefined%
  \makeatother%
  \begin{picture}(1,0.47354432)%
    \lineheight{1}%
    \setlength\tabcolsep{0pt}%
    \put(0,0){\includegraphics[width=\unitlength,page=1]{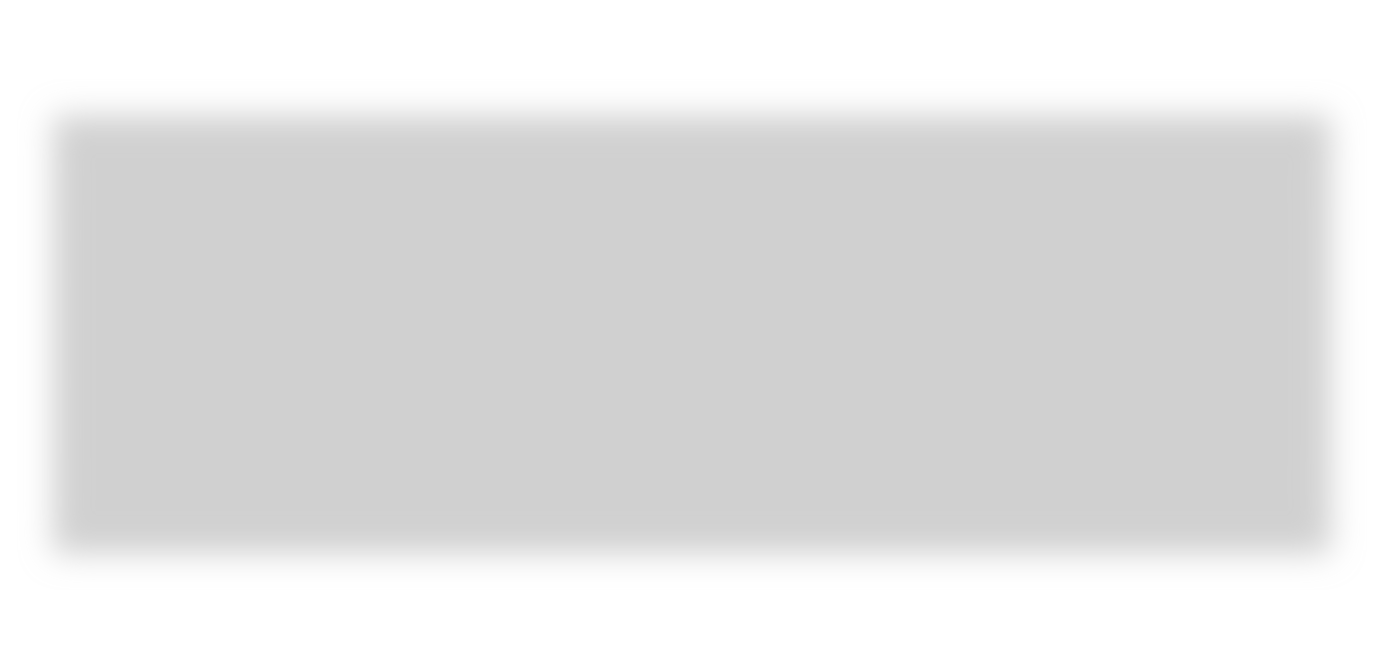}}%
    \put(0.76191111,0.3562397){\color[rgb]{0,0,0}\makebox(0,0)[lt]{\lineheight{1.25}\smash{\begin{tabular}[t]{l}$j^{(ext)}$\end{tabular}}}}%
    \put(0.08858347,0.37567946){\color[rgb]{0,0,0}\makebox(0,0)[lt]{\lineheight{1.25}\smash{\begin{tabular}[t]{l}$j^{(ext)}$\end{tabular}}}}%
    \put(0.07113966,0.0967765){\color[rgb]{0,0,0}\makebox(0,0)[lt]{\lineheight{1.25}\smash{\begin{tabular}[t]{l}$j^{(ext)}$\end{tabular}}}}%
    \put(0,0){\includegraphics[width=\unitlength,page=2]{Turbulence2.pdf}}%
    \put(0.89900002,0.21279342){\color[rgb]{0,0,0}\makebox(0,0)[lt]{\lineheight{1.25}\smash{\begin{tabular}[t]{l}$j^{(ext)}$\end{tabular}}}}%
    \put(0,0){\includegraphics[width=\unitlength,page=3]{Turbulence2.pdf}}%
    \put(0.30662744,0.43053632){\color[rgb]{0,0,0}\makebox(0,0)[lt]{\lineheight{1.25}\smash{\begin{tabular}[t]{l}\huge{$\mathbf{u}$}\end{tabular}}}}%
    \put(0.84410954,0.3132519){\color[rgb]{1,0.00392157,0}\makebox(0,0)[lt]{\lineheight{1.25}\smash{\begin{tabular}[t]{l}$\delta B$\end{tabular}}}}%
    \put(0.86589938,0.16741395){\color[rgb]{0,0.16470588,1}\makebox(0,0)[lt]{\lineheight{1.25}\smash{\begin{tabular}[t]{l}$\delta B$\end{tabular}}}}%
    \put(0,0){\includegraphics[width=\unitlength,page=4]{Turbulence2.pdf}}%
    \put(0.48417882,0.24847611){\color[rgb]{0,0,0}\makebox(0,0)[lt]{\lineheight{1.25}\smash{\begin{tabular}[t]{l}\large{$\mathbf{V}$}\end{tabular}}}}%
  \end{picture}%
\endgroup%

}

\newcommand{\infg}[1]{%
    \def\svgwidth{\columnwidth}
\begingroup%
  \makeatletter%
  \providecommand\color[2][]{%
    \errmessage{(Inkscape) Color is used for the text in Inkscape, but the package 'color.sty' is not loaded}%
    \renewcommand\color[2][]{}%
  }%
  \providecommand\transparent[1]{%
    \errmessage{(Inkscape) Transparency is used (non-zero) for the text in Inkscape, but the package 'transparent.sty' is not loaded}%
    \renewcommand\transparent[1]{}%
  }%
  \providecommand\rotatebox[2]{#2}%
  \newcommand*\fsize{\dimexpr\f@size pt\relax}%
  \newcommand*\lineheight[1]{\fontsize{\fsize}{#1\fsize}\selectfont}%
  \ifx\svgwidth\undefined%
    \setlength{\unitlength}{273.71997503bp}%
    \ifx\svgscale\undefined%
      \relax%
    \else%
      \setlength{\unitlength}{\unitlength * \real{\svgscale}}%
    \fi%
  \else%
    \setlength{\unitlength}{\svgwidth}%
  \fi%
  \global\let\svgwidth\undefined%
  \global\let\svgscale\undefined%
  \makeatother%
  \begin{picture}(1,0.78579315)%
    \lineheight{1}%
    \setlength\tabcolsep{0pt}%
    \put(0,0){\includegraphics[width=\unitlength,page=1]{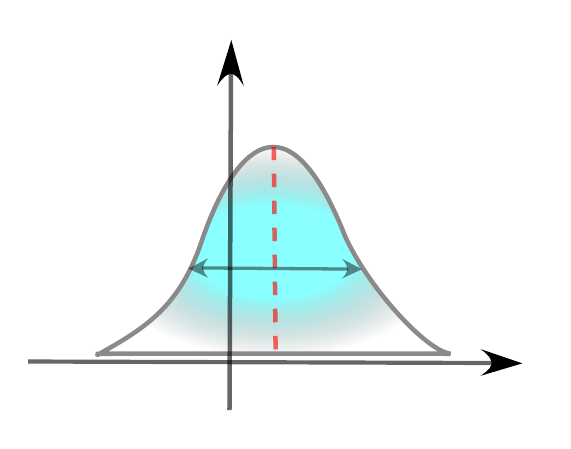}}%
    \put(0.42889563,0.69136106){\color[rgb]{0,0,0}\makebox(0,0)[lt]{\lineheight{1.25}\smash{\begin{tabular}[t]{l}\Huge{$f_{0}(v)$}\end{tabular}}}}%
    \put(0.47434401,0.0901518){\color[rgb]{0,0,0}\makebox(0,0)[lt]{\lineheight{1.25}\smash{\begin{tabular}[t]{l}\Huge{$u$}\end{tabular}}}}%
    \put(0.64179723,0.33835973){\color[rgb]{0,0,0}\makebox(0,0)[lt]{\lineheight{1.25}\smash{\begin{tabular}[t]{l}\Huge{$2v_{T}$}\end{tabular}}}}%
    \put(0.87850986,0.09194671){\color[rgb]{0,0,0}\makebox(0,0)[lt]{\lineheight{1.25}\smash{\begin{tabular}[t]{l}\Huge{$v$}\end{tabular}}}}%
    \put(0.35334543,0.09656695){\color[rgb]{0,0,0}\makebox(0,0)[lt]{\lineheight{1.25}\smash{\begin{tabular}[t]{l}\Huge{$\mathbf{0}$}\end{tabular}}}}%
  \end{picture}%
\endgroup%

}
\title{Skin_Effects_in_Magnetic_Turbulence_Diffusion}
\author{nick.emelyanov2014 }
\date{September 2025}

\begin{document}
\begin{center}
	
	{\large \textbf{The skin effect in anomalous transport of charged particles in plasma with a microturbulent magnetic field. I. Isotropic plasma 
    }}\\
	\medskip
	{ \textbf{N.\:A.~Emelyanov$^{1,2}$, Vl.\:V.~Kocharovsky$^{1}$}}\\
	\medskip
	\textit{$^{1}$A.V. Gaponov-Grekhov Institute of Applied Physics of the Russian Academy of Sciences,\\ Ulyanov str. 46, Nizhny Novgorod, 603950 Russia\\}
    \textit{$^{2}$ Pulkovo Astronomical Observatory, Russian Academy of Sciences,\\ Pulkovo chaussee 65, Saint Petersburg, 196140 Russia\\ }

	\textit{n.emelyanov@ipfran.ru}
	
\end{center}
%%%%%%%%%%%%%%%%%%%% ABSTRACT %%%%%%%%%%%%% %%%%%%%%%%%%%%%%%%%%%%%%%%%%%%%%%%%%%%%%%%%
\begin{abstract}
\label{sec0}
\par The influence of electromagnetic skin effect on anomalous charged particle transport in dense, non-relativistic, collisionless plasma with a small-scale turbulent magnetic field was investigated using quasi-linear kinetic equations, through both analytical and numerical methods. Analytical expressions for the  diffusion tensor components in the Fokker–Planck equation that take this effect into account have been found. The equation was solved numerically in the case of magnetostatic turbulence. It has been demonstrated that the skin  effect increases the mean free path of particles in turbulent plasma, thereby reducing its anomalous resistance. It also leads to anisotropy in particle scattering, resulting in anisotropy in their stationary velocity distribution, which increases as the screening parameter grows. Approximate analytical formulas for the effective mobility of charged particles and the electric conductivity of plasma with isotropic magnetostatic turbulence have been obtained.
\end{abstract}
%%%%%%%%%%%%%%%%%% INTRODUCTION %%%%%%%%%%% %%%%%%%%%%%%%%%%%%%%%%%%%%%%%%%%%%%%%%%%%%%
\section{Introduction}
\label{sec1}
\par Studying and interpretation a variety of phenomena in laboratory and space plasmas usually requires to determine their kinetic characteristics, such as diffusion coefficients, viscosity and electrical or thermal conductivity \cite{bobrovConductivityDiffusionCoefficients2019,lafleurFrictionStronglyMagnetized2020,joseGeneralizedBoltzmannKinetic2020,dongImpactMagneticField2022,dongCollisionTermUniformly2023a,waybrightEffectsElectronViscosity2024,leeInvestigationResonantLayer2025,numataQuasilinearDissipationEstimates2025}. However, plasma, especially cosmic plasma, is often non-collisional: on scales of interest, the main role in the kinetics of charged particles is played not by their pair collisions, but by their interaction with turbulent fluctuations of electric and magnetic fields \cite{rostokerFluctuationsPlasma1961,dupreeKineticTheoryPlasma1963,rostokerKineticTheoryPlasma1965,ichimaruRelaxationProcessesPlasmas1970,newmanGeneralizationEquationsGoverning1973,chuMagnetostaticModeCrossField1978,okudaPlasmaDiffusionDue1979,mynickParticleStochasticityDue1980,subediChargedParticleDiffusion2017,yoonKineticInstabilitiesSolar2017,shalchiPerpendicularTransportEnergetic2020,zhangCosmicRayDiffusion2024,islikerTransportParticlesStrongly2025}. In this case, the corresponding coefficients may differ by many orders of magnitude from the \say{classical} values in Coulomb scattering, and the plasma parameters are referred to as anomalous. 
\par  Research on the transport properties of turbulent plasma is of great importance for solving a number of problems in controlled thermonuclear fusion \cite{rechesterElectronHeatTransport1978,okudaPlasmaDiffusionDue1979,mynickGeneralizedBananadriftTransport1986,atzeniInertialConfinementFusion2013,robinsonTheoryFastElectron2014,dongImpactMagneticField2022, vogmanParameterizedAnomalousTransport2025}, as well as a wide range of tasks in space plasma physics, such as explaining the origin of high-energy cosmic rays \cite{ginzburgOriginCosmicRays1964,jokipiiCosmicRayPropagationCharged1966,longairHighEnergyAstrophysics1992,schlickeiserQuasilinearTheoryCosmic1998,medvedevGenerationMagneticFields1999,casseTransportCosmicRays2001a,plotnikovParticleTransportIntense2011,shalchiPerpendicularTransportEnergetic2020,zhangCosmicRayDiffusion2024}  or the occurrence of anomalous conductivity in current layers during solar flares and magnetospheric substorms \cite{lazarianTurbulenceMagneticReconnection2012,kontarTURBULENTPITCHANGLESCATTERING2013, karimabadiMagneticReconnectionPresence2013,bianSUPPRESSIONPARALLELTRANSPORT2016,kontarTurbulentKineticEnergy2017,mussetDiffusiveTransportEnergetic2018,ergunParticleAccelerationStrong2020,lazarian3DTurbulentReconnection2020,Vicentin_2025}. A huge number of works and monographs are devoted to the study of anomalous transport of charged particles (see., e.g., \cite{ginzburgOriginCosmicRays1964, cytovicIntroductionTheoryPlasma1972,davidsonMethodsNonlinearPlasma1972, akhiezerPlasmaElectrodynamics21975}).  
\par The main focus of this study is  often on the interaction of charged particles with the  electric field fluctuations  \cite{rostokerKineticEquationConstant1960b, rostokerFluctuationsPlasma1961,ichimaruRelaxationProcessesPlasmas1970,cytovicIntroductionTheoryPlasma1972,davidsonMethodsNonlinearPlasma1972,hassamConvectiveCellsTransport1979,dongCollisionTermUniformly2023}. However, in some cases, the separation of charges in plasma is negligible, or turbulent magnetic fluctuations are excited significantly above the thermal level due to various hydrodynamic and kinetic instabilities (see, e.g., \cite{bellTurbulentAmplificationMagnetic2004,yoonCollisionalRelaxationBiMaxwellian2016,zhouTurbulentMixingTransition2019,matthaeusTurbulenceSpacePlasmas2021,shaabanNewLowbetaRegime2021,zhouSpontaneousMagnetizationCollisionless2022,takabeTheoryMagneticTurbulence2023,zhouMagnetogenesisCollisionlessPlasma2024,zhangCosmicRayDiffusion2024}), there may be a regime in which particle scattering in a turbulent magnetic field plays a more important role. Examples include the interaction of cosmic rays with interplanetary or galactic magnetic fields and the acceleration of particles in collisionless shock waves \cite{schlickeiserQuasilinearTheoryCosmic1998,medvedevTheoryJitterRadiation2000,keenanParticleTransportRadiation2013,bellInteractionCosmicRays2005,bykovCosmicRayCurrent2011,bykovFundamentalsCollisionlessShocks2011,subediChargedParticleDiffusion2017,zhangCosmicRayDiffusion2024}.
\par However, in these cases, the scattering of individual particles is usually considered as scattering in a random field supported by some external sources in a vacuum, and the influence of collective effects, including the particles' own fields, is neglected (see., e.g., \cite{schlickeiserQuasilinearTheoryCosmic1998,shalchiPlasmaparticleInteractionStrong2009,plotnikovParticleTransportIntense2011,subediChargedParticleDiffusion2017}). This is acceptable for particles moving at significantly greater speeds than thermal velocities, or for highly rarefied cosmic plasma. However, as will be demonstrated below, for sufficiently dense beams of charged particles, moving at speeds equal to or less than thermal velocities, the effect of their own screening can be significant. This can lead to anisotropy in the transport properties of the plasma and a decrease in its anomalous resistance.   
\par This paper presents an analytical and numerical study of the influence of electromagnetic shielding on the transport properties of a dense homogeneous non-relativistic collisionless plasma flow in the presence of a microturbulent magnetic field. The quasi-linear kinetic equation method is used to describe the transport of charged particles. It is correct, if scattering on magnetic fluctuations is weak, and the plasma is unmagnetized.
\par  The task at hand is considered in its most general formulation, in which interaction with the magnetic turbulence of a selected group of charged particles generating electric current in plasma causes relaxation of their average directed motion and, accordingly, the appearance of anomalous resistance. In real conditions, the role of scattered particles can be performed by either one of the plasma components, such as electrons or ions, or by injected particle beams having an average velocity relative to the given turbulent field sources. 
\par This paper contains the following sections. Section \ref{sec2} presents the derivation of the Fokker-Planck (FP) equation for a dense beam of charged particles in turbulent plasma, and the corresponding components of the diffusion tensor in velocity space are found in the quasi-linear approximation. Section \ref{sec3} considers the case where the turbulent magnetic field is stationary and isotropic, and expressions are obtained for the characteristic screening length and effective diffusion coefficient. Section \ref{sec4} is devoted to the numerical solution of the resulting FP equation and a discussion of the effect of screening on the emergence of anisotropy in the particle distribution function and the increase in their mean free path. Section \ref{sec5} consists of conclusions and a discussion of open questions.
%%%%%%%%%%%%%%%%%%%%%%%%%%%%%%%
\section{Particle diffusion tensor in velocity space}
\label{sec2}
\par Let us consider the propagation of a non-relativistic flow (beam) of charged particles, such as electrons, in an isotropic homogeneous collisionless plasma with an excited small-scale turbulent magnetic field, which is maintained by some external currents, for example, currents of charged particles of a different type or a separate fraction of the same particles. At high beam concentrations, external sources induce significant polarization currents in the beam, which, in turn, lead to changes in the effective scattering fields. Thus, the particle transport problem is, strictly speaking, non-linear, since the scattering electromagnetic fields must be consistent with the solution of the kinetic equation for the averaged distribution function that determines the dielectric properties of the beam plasma. This problem is similar to that which arises when describing particle scattering in a Coulomb field using the Balescu-Lenard collision integral (see., e.g., \cite{lenardBogoliubovsKineticEquation1960,balescuIrreversibleProcessesIonized1960,dongDerivationMagnetizedBalescuLenardGuernsey2017,dongCollisionTermUniformly2023a}).   
\par The turbulence is assumed to be sufficiently weak, so that when the beam particles scatter, the deviation of their direction of motion at a distance of the order of the characteristic correlation length of the magnetic field $\lambda_{cor}$ occurs at a small angle, and the number of magnetized particles is negligible. In other words, we will assume that the condition $l_{c}\gg \bar{r}_{H}\gg \lambda_{cor}$ is satisfied, where $l_{c}$ is the mean free path of a particle determined by turbulence, $\bar{r}_ {H}$, is the thermal gyroradius of particles in an average turbulent magnetic field $\bar{B}_{T}=\sqrt{\langle B^2 \rangle}$. Under these conditions, the change in particle velocities is diffusive in nature and is described by the FP equation \cite{hubbardFrictionDiffusionCoefficients1961,landauCourseTheoreticalPhysics101981,riskenFokkerPlanckEquationMethods1996,dongCollisionTermUniformly2023a}. To find a specific type of coefficients included in this equation, we use the kinetic description of Vlasov-Maxwell in a quasi-linear approximation \cite{cytovicIntroductionTheoryPlasma1972,davidsonMethodsNonlinearPlasma1972, schlickeiserQuasilinearTheoryCosmic1998,dongCollisionTermUniformly2023} and briefly go through the well-known scheme of their derivation.
\par Let us write down the Liouville equation for the velocity distribution function of a particle beam $F(t,\mathbf{r},\mathbf{v})$ with mass $m$ and charge $q$, moving in a random electromagnetic field defined by the electric field and magnetic field vectors,  $\mathbf{E}$ and $\mathbf{B}$ respectively. Let us assume that the average vectors of these fields are zero: $\langle\mathbf{E}\rangle$, $\langle\mathbf{B}\rangle=0$. The possible average value of the external electric field $\mathbf{E}_d$ supporting the flow of direct current in plasma, for simplicity, we will also consider it to be zero (until Section \ref{sec4}), since its consideration (at a sufficiently small value) only leads to the appearance of a trivial additional term in the equation for the averaged distribution function. In this case, the equation is
\begin{align}
    \label{Kinetic_Eq1}
    \frac{\partial F}{\partial t}+\mathbf{v}\cdot \frac{\partial F}{\partial \mathbf{r}} +\frac{q}{m} \Big\{\mathbf{E}+\frac{1}{c}(\mathbf{v}\times\mathbf{B}) \Big\} \cdot\frac{\partial F}{\partial \mathbf{v}}=0.
\end{align}
\par Ignoring electrostatic disturbances, including Langmuir and ion-sound turbulence, we express the vectors of electric field strength and magnetic induction,  
\begin{align}
    \label{Field_expressions}
     \mathbf{E}=-\frac{1}{c}\frac{\partial \mathbf{A}}{\partial t},\ \ \ 
     \mathbf{B}=\nabla\times \mathbf{A},
\end{align}
through the vector potential $\mathbf{A}$, which is the solution to the wave equation
\begin{align}
    \label{A_wave_equation}
    \nabla^2\mathbf{A}-\frac{1}{c^2}\frac{\partial}{\partial t}\hat{\varepsilon}_{t}\frac{\partial\mathbf{A}}{\partial t}=-\frac{4\pi}{c}\mathbf{j}^{(ext)}.
\end{align}
Here, $\hat{\varepsilon}_{t}$ is the dielectric permittivity operator, which takes into account the influence of polarization currents in a homogeneous stationary plasma, $\mathbf{j}^{(ext)}$ is the electric current density of external sources, and $c$ is the speed of light in a vacuum. In expressions (\ref{Field_expressions}), (\ref{A_wave_equation}) and further, Coulomb calibration of the vector potential, $\nabla\cdot\mathbf{A}=0$, is used.
\par The kinetic equation (\ref{Kinetic_Eq1}) can be rewritten as follows:
\begin{align}
    \label{Kinetic_eq2}
    \frac{\partial F}{\partial t}+\mathbf{v}\cdot \frac{\partial F}{\partial \mathbf{r}}+ \hat{D}[\mathbf{A}]\cdot \frac{\partial F}{\partial \mathbf{v}}=0,
\end{align}
where, for brevity, a linear differential operator is introduced
\begin{align}
    \label{Diff_operator}
     \hat{D}[\mathbf{A}]=-\frac{q}{mc}\Big\{\frac{\partial \mathbf{A}}{\partial t}-\mathbf{v}\times(\nabla\times\mathbf{A}) \Big\}.
\end{align}
\par As turbulence is considered to be weak, the solution of the equation (\ref{Kinetic_Eq1}) can be found with use of perturbation theory as a series in terms of the small energy of turbulent fluctuations,   
\begin{align}
    \label{Perturb_series}
    F=f_{0}+f_{1}+f_{2}+...
\end{align}
\par Following the standard procedure of quasi-linear theory \cite{schlickeiserQuasilinearTheoryCosmic1998,dongCollisionTermUniformly2023},  the distribution function $F$ in the specified form should be substituted in equation (\ref{Kinetic_Eq1}). Averaging over the statistical distribution of the value $\mathbf{A}$ (symbol $\langle...\rangle$) and omitting terms of higher order of smallness than the first, one can obtain the following system of equations:
\begin{align}
    \label{Aver_kin_eq}
    \frac{\partial f_{0}}{\partial t}+\mathbf{v}\cdot \frac{\partial f_{0}}{\partial \mathbf{r}} + \langle\hat{D}[\mathbf{A}]\cdot \frac{\partial f_{1}}{\partial \mathbf{v}}\rangle=0,\\
     \label{perturb_dist_f}
    \frac{\partial f_{1}}{\partial t}+\mathbf{v}\cdot \frac{\partial f_{1}}{\partial \mathbf{r}}+ \hat{D}[\mathbf{A}] \cdot\frac{\partial f_{0}}{\partial \mathbf{v}}=0.
\end{align}
\par Considering thermal fluctuations to be sufficiently weak, we restrict our search to the forced solution of equation (\ref{perturb_dist_f}). After the Fourier transform an expression for the particle distribution function perturbation $f_{1}$ as a function of frequency and wave vector is written (summation over repeated vector and tensor indices is assumed):
\begin{align}
    \label{Sol_pert_kin_eq}
     f_{1}(\omega,\mathbf{k})=-i\hat{D}_{\alpha\beta}\frac{\partial f_{0}}{\partial v_{\alpha}}\frac{A_{\beta}(\omega,\mathbf{k})}{\omega-\mathbf{k}\cdot\mathbf{v}}.
\end{align}
Here, $\hat{D}_{\alpha\beta}$ is the matrix of the operator  $\hat{D}$ in Fourier representation, whose components are explicitly written as follows: 
\begin{align}
    \label{Spec_Diff_operator}
    \hat{D}_{\alpha\beta}(\omega,\mathbf{k})=\frac{iq}{mc}\Big\{(\omega-\mathbf{k}\cdot\mathbf{v})\delta_{\alpha\beta}+k_{\alpha}v_{\beta} \Big\}.
\end{align}
The Fourier transform $\mathbf{A}(\omega,\mathbf{k})$ is the corresponding solution of equation (\ref{A_wave_equation}) in spectral form:
\begin{align}
    \label{solution_wave_eq}
    \mathbf{A}(\omega,\mathbf{k})=\frac{4\pi}{c}\frac{\mathbf{j}^{(ext)}(\omega,\mathbf{k})}{k^2\Lambda(\omega,\mathbf{k})}.
\end{align}
In relation (\ref{solution_wave_eq}), there is a factor $\Lambda$, expressed through the transverse dielectric permittivity of plasma $\varepsilon_{t}$, which is responsible for the effect of screening \footnote{In the quasi-linear approximation, it is assumed that the value of the dielectric permittivity of the beam plasma $\varepsilon_ {t}$ is determined by the undisturbed  particle velocity distribution function $f_{0}$, which imposes restrictions on the applicability of relation (\ref{Sol_pert_kin_eq}). In fact, to use the method described, it is necessary that all characteristic spatial and temporal scales of variation of the function $f_{0}$ be much larger than the corresponding scales of turbulent fluctuations.}:
\begin{align}
    \label{Lambda_factor}
    \Lambda(\omega,\mathbf{k})=1-\frac{\omega^2}{k^2c^2}\varepsilon_{t}(\omega,\mathbf{k}).
\end{align}
In the case where there are no external sources, $\mathbf{j}^{(ext)}=0$, the equality of expression (\ref{Lambda_factor}) to zero defines the dispersion relation for transverse normal waves in undisturbed plasma. However, in the following, we will only be interested in the forced solution, for which $\Lambda\ne0$.
\par After some simple transformations using the expression found (\ref{Sol_pert_kin_eq}), the equation for the averaged distribution function (\ref{Aver_kin_eq}) can be rewritten as
\begin{align}
    \label{Fokker_Plank}
    \frac{\partial f_{0}}{\partial t}+\mathbf{v}\cdot \frac{\partial f_{0}}{\partial \mathbf{r}}=\frac{\partial }{\partial v_{\alpha}}\hat{\mu}_{\alpha\beta}\frac{\partial f_{0}}{\partial v_{\beta}}.
\end{align}
This expression has the form of a FP equation, where the tensor (matrix) of diffusion coefficients in velocity space is determined by the following formula (the symbol $^*$ denotes complex conjugation, the upper index $(ext)$ for the current density components is omitted here and below): 
\begin{align}
    \label{Diffus_coeff_0}
    \hat{\mu}_{\alpha\beta}=i\Big(\frac{4\pi}{c} \Big)^2\int\int \frac{\hat{D}_{\alpha\gamma}^{*} \hat{D}_{\beta\gamma'}}{\omega-\mathbf{k}\cdot\mathbf{v}}\frac{\langle j_{\gamma}^*(\omega',\mathbf{k}')j_{\gamma'}(\omega,\mathbf{k})\rangle}{k'^2k^2\Lambda^*(\omega',\mathbf{k}')\Lambda(\omega,\mathbf{k})}\exp{\big(-i(\omega-\omega')t+i(\mathbf{k}-\mathbf{k}')\cdot\mathbf{r}\big)}\frac{d^3k  d^3k'd\omega d\omega'}{(2\pi)^8}.
\end{align}
\par In assumption of stationary and homogeneous turbulence,  the well-known relationship between the correlation function and the spectral density of fluctuations can be used (see, for example, \cite{akhiezerPlasmaElectrodynamics21975})
\begin{align}
    \label{Correlation_func}
    \langle j_{\gamma}^*(\omega',\mathbf{k}')j_{\gamma'}(\omega,\mathbf{k})\rangle=(2\pi)^4\bar{j}_{\gamma\gamma'}^2(\omega,\mathbf{k})\delta(\omega-\omega')\delta(\mathbf{k}-\mathbf{k}').
\end{align}
Here, $\bar{j}^{2}_{\gamma'\gamma}$ is the spectral distribution of fluctuations in the electric current density of external sources, and $\delta(x)$ is the Dirac function. Using relation (\ref{Correlation_func}), one can find the diffusion coefficient matrix in the following form:
\begin{align}
    \label{Diffus_coeff}
    \hat{\mu}_{\alpha\beta}=i\Big(\frac{4\pi}{c} \Big)^2\int \frac{\hat{D}_{\alpha\gamma}^{*} \hat{D}_{\beta\gamma'}}{\omega-\mathbf{k}\cdot\mathbf{v}}\frac{\bar{j}_{\gamma\gamma'}^2(\omega,\mathbf{k})}{|\Lambda(\omega,\mathbf{k})|^2}\frac{d^3k  d\omega}{(2\pi k)^4}.
\end{align}
\par The expression (\ref{Diffus_coeff}) found for the coefficients of the FP equation describing the transport of charged particles in a random electromagnetic field is the most general that can be obtained for stationary homogeneous turbulence in the absence of two-particle collisions. If   the distribution of turbulent currents is considered to be isotropic, i.e. satisfying the relation
\begin{align}  
    \label{Isotrop_turb}
    \bar{j}_{\gamma\gamma'}^2(\omega,\mathbf{k})=\frac{1}{3}\bar{j}^2(\omega,k)\delta_{\gamma\gamma'},
\end{align}
then, after simple transformations, the expression (\ref{Diffus_coeff}) is
\begin{align}
    \label{D_coeff_isotrop}
    \hat{\mu}_{\alpha\beta}=\frac{i}{3}\Big(\frac{4\pi q}{mc^2} \Big)^2\int \frac{v^2k_{\alpha}k_{\beta}}{\omega-\mathbf{k}\cdot\mathbf{v}}\frac{\bar{j}^2(\omega,k)}{|\Lambda(\omega,k)|^2}\frac{d^3k  d\omega}{(2\pi k)^4}.
\end{align}
\par Following the rule for handling singularities in the denominator of integrand of the Fourier transform \cite{landauCourseTheoreticalPhysics1965} (the symbol $P$ denotes integration in the sense of the principal value of Cauchy)
\begin{align}
    \label{Landau_pole}
    \frac{1}{\omega-\mathbf{k\cdot\mathbf{v}}+i0}\longrightarrow P\frac{1}{\omega-\mathbf{k\cdot\mathbf{v}}}-i\pi\delta(\omega-\mathbf{k\cdot\mathbf{v}}),
\end{align}
the following tensor of particle diffusion coefficients in velocity space appears: 
\begin{align}
    \label{Diffus_tensor}
    \hat{\mu}_{\alpha\beta}=\frac{2\pi}{3}\Big(\frac{4\pi q}{mc^2} \Big)^2\frac{\mathcal{E}}{m}\int \frac{k_{\alpha}k_{\beta}}{k^2}\frac{\bar{j}^2(\omega,k)}{|\Lambda(\omega,k)|^2}\frac{\delta(\omega-\mathbf{k\cdot\mathbf{v}})}{k^2}\frac{d^3k  d\omega}{(2\pi)^4},
\end{align}
where $\mathcal{E}=mv^2/2$ is the kinetic energy of the particles. It is easy to show that formula (\ref{Diffus_tensor}) can be presented in a more intuitive form 
\begin{align}
    \label{Diffus_tensor_struct}
    \hat{\mu}_{\alpha\beta}=\mu_{\perp}\delta_{\alpha\beta}+(\mu_{||}-\mu_{\perp})\frac{v_{\alpha}v_{\beta}}{v^2}.
\end{align}
Here, the longitudinal and transverse diffusion coefficients are equal to, respectively \footnote{In reality, integration over the wave number $k$ must be carried out up to a certain maximum value $k_{max}$, determined by the condition of applicability of the continuous medium approximation in relation (\ref{solution_wave_eq}). However, in this work, it is assumed that the characteristic correlation lengths of turbulence are much larger than the specified scale, and therefore the integration can be extended to infinity.}: 
\begin{align}
    \label{Long_diffus_coeff}
    \mu_{||}(\mathcal{E})=\frac{1}{3}\Big(\frac{4\pi q}{mc^2} \Big)^2\sqrt{\frac{\mathcal{E}}{2m}}\int_{0}^{+\infty} \Big(1-\frac{m\omega^2}{2\mathcal{E}k^2}\Big)^{3/2}\frac{\bar{j}^2(\omega,k)}{|\Lambda(\omega,k)|^2}\frac{dk  d\omega}{k(2\pi)^2},
\end{align}
\begin{align}
    \label{Perp_diffus_coeff}
    \mu_{\perp}(\mathcal{E})=\frac{1}{6}\Big(\frac{4\pi q}{mc^2} \Big)^2\sqrt{\frac{\mathcal{E}}{2m}}\int_{0}^{+\infty} \frac{m\omega^2}{2\mathcal{E}k^2}\Big(1-\frac{m\omega^2}{2\mathcal{E}k^2}\Big)^{1/2}\frac{\bar{j}^2(\omega,k)}{|\Lambda(\omega,k)|^2}\frac{dk  d\omega}{k(2\pi)^2}.
\end{align}
\par The expressions found (\ref{Long_diffus_coeff}) and (\ref{Perp_diffus_coeff}) differ from the standard diffusion coefficients in particle scattering theory (in particular, cosmic rays) on magnetic turbulence (see, for example, \cite{casseTransportCosmicRays2001a,keenanParticleTransportRadiation2013,subediChargedParticleDiffusion2017}) by the presence of an additional integrand factor, $1/|\Lambda|^2$, responsible for the skin effect. The peculiarities of plasma transport properties arising from this factor are analysed in the following sections.
\par From the relation (\ref{Diffus_tensor_struct}), the physical meaning of the obtained formulas is clearly visible: the longitudinal coefficient $\mu_{||}$ describes the diffusion of particle energy, while the transverse coefficient $\mu_{\perp}$ describes the diffusion of angles. Substituting the found expressions into (\ref{Fokker_Plank}), we obtain a closed equation for the averaged distribution function $f_{0}$ describing the transport of particles in plasma with a microturbulent magnetic field. However, in formulas (\ref{Long_diffus_coeff}) and (\ref{Perp_diffus_coeff}), the spectral density of current fluctuations $\bar{j}^{2}(\omega,k)$ remains undefined and has been considered known and given by external sources when deriving these relations. In fact, it must be derived from experimental data or model assumptions about the properties of the turbulent field. 
%%%%%%%%%%%%%%%%%%% SECTION_3 %%%%%%%%%%%%% %%%%%%%%%%%%%%%%%%%%%%%%%%%%%%%%%%%%%%%%%%%
\section{Scattering of particles by magnetostatic perturbations} 
\label{sec3}
\par If the characteristic pulsation time of the magnetic field $\tau_{B} $ is sufficiently large at the times of interest (i.e., $\tau_B\gg\tau_c$, where $\tau_c$ is the average effective free path time of particles in turbulent plasma without significant changes in their direction or magnitude), then such a field can be considered quasi-stationary and treated as a time-fixed spatial structure. In a number of works (see, for example, \cite{chuMagnetostaticModeCrossField1978, okudaPlasmaDiffusionDue1979,pavlenkoNonlinearMagneticIslands1981,plotnikovParticleTransportIntense2011}) based on a simple hydrodynamic approach, it was shown that such magnetostatic perturbations can exist in plasma for a sufficiently long time and cause strong particle scattering, more significant than the scattering associated with Coulomb collisions, especially if the level of magnetic fluctuations is superthermal. As a result, anomalous particle diffusion corresponding to the Bohm-like regime and anomalous electron thermal conductivity across an external uniform magnetic field arise in the plasma \cite{chuMagnetostaticModeCrossField1978, pavlenkoNonlinearMagneticIslands1981}.
Below, using the results of kinetic theory presented in the previous section, the diffusion coefficients of the FP equation are derived and an estimate is given for the anomalous conductivity of the beam plasma, determined by the transport of particles in an isotropic microturbulent magnetic field, under conditions of weak particle magnetization and zero or negligible average magnetic field induction vector. 
\par In cases where the spectral distribution of electric current density fluctuations corresponds to magnetostatic disturbances, the following equality holds true:
\begin{align}
    \label{Stat_turb_2}    \bar{j}^2(\omega,k)=2\pi\bar{j}^2(k)\delta(\omega).
\end{align}
\par Furthermore,  the beam of scattered particles is assumed to have an initial velocity distribution close to Maxwellian and moves relative to stationary turbulent sources at an average velocity $\mathbf{u}$. Then the diffusion coefficient tensor (\ref{Diffus_tensor}) takes the form:
\begin{align}
    \label{Diffus_tensor_convect_2}    \hat{\mu}_{\alpha\beta}=\frac{2\pi}{3}\Big(\frac{4\pi q}{mc^2} \Big)^2\frac{\mathcal{E}}{m}\int \frac{k_{\alpha}k_{\beta}}{k^2}\frac{\bar{j}^2(k)}{|\Lambda(\mathbf{k\cdot u},k)|^2}\frac{\delta(\mathbf{k}\cdot\mathbf{v})}{k^2}\frac{d^3k  }{(2\pi)^3}.
\end{align}
\par
\begin{figure}[t]
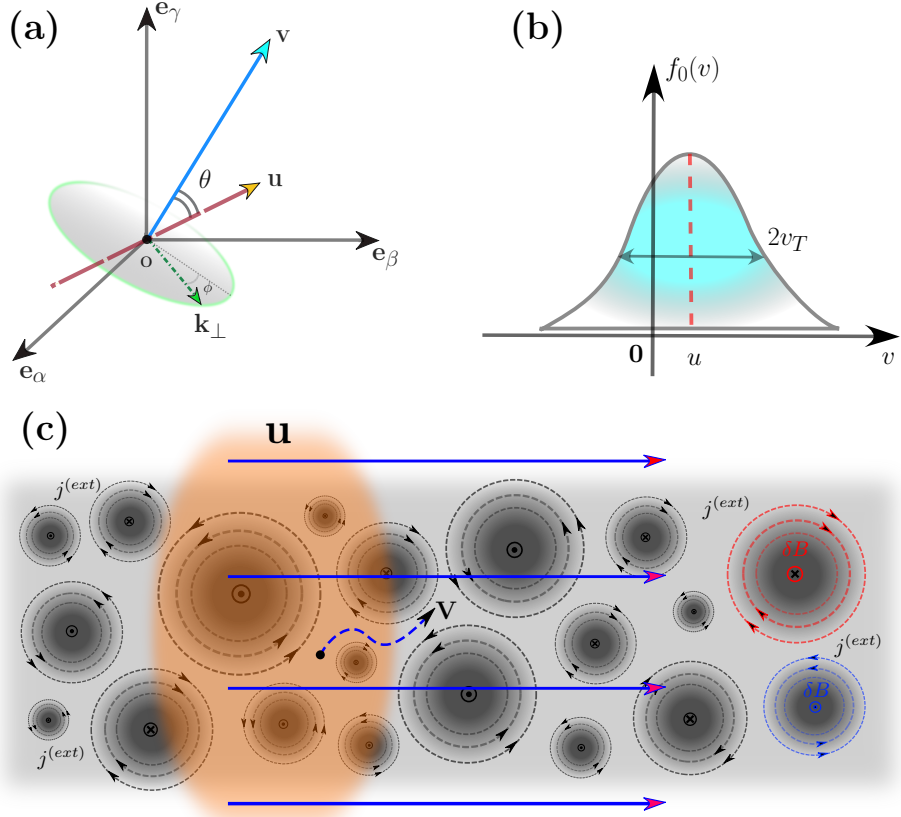
  %%%%%%%%%%%%%%%%%% FIGURE 1
\centerline{\Large \bf     
\hspace{0.12 \textwidth}  \color{black}{(a)}
\hspace{0.35 \textwidth}  \color{black}{(b)}
   \hfill}
   \vspace{-0.06\textwidth}
 \centering
    %\vspace{-0.31\textwidth} 
    %\includegraphics[width=0.5\linewidth]{Pictures/footpoint_new.pdf} 
    \scalebox{0.4}{\incfig{Coordinat_system_mod}}
    \scalebox{0.4}{\infg{Func_distrib}}
\centerline{\Large \bf     
\hspace{0.13 \textwidth}  \color{black}{(c)}
   \hfill}
   \vspace{-0.06\textwidth}
    
    \scalebox{0.8}{\incfg{Turbulence2}}
    
    \caption{ a) Schematic representation of a laboratory reference frame with guide axes $\mathbf{e}_{\alpha,\beta,\gamma}$ and the direction of motion of a single particle with velocity $\mathbf{v}$ and a particle beam with drift velocity $\mathbf{u}$. The shaded area shows the resonance plane of the wave vectors of the turbulent magnetic field $\mathbf{k}_{\perp}$ that contribute non-zero to particle scattering. b) Schematic representation of the averaged distribution function of scattered beam particles by velocity. The average values of the drift and double thermal velocities ($u$ and $v_{T}$) are marked. c) Symbolic two-dimensional picture of the propagation of a particle beam in plasma with a turbulent magnetic field.}
    \label{fig:system}
\end{figure}

 From the expression found (\ref{Diffus_tensor_convect_2}), it can be seen that only the harmonics of the magnetic field with wave vectors that are orthogonal to the direction of its motion (see Fig. \ref{fig:system}a) and which, for brevity, will be denoted as $\mathbf{k}_{\perp}$. When integrating over all directions of wave vectors, it is convenient to count the polar angle from the direction of the particle velocity $\mathbf{v}$, and the azimuthal angle $\phi$ from the direction of an arbitrary axis in the orthogonal plane (the plane of wave vectors $\mathbf{k}_{\perp}$ shown in Fig. \ref{fig:system}a). Thus, we arrive at the following relationship:
\begin{align}
    \label{Diffus_tensor_conv2}    \hat{\mu}_{\alpha\beta}=\frac{1}{3}\Big(\frac{4\pi q}{mc^2} \Big)^2\sqrt\frac{\mathcal{E}}{2m}\int \frac{k_{\perp\alpha}k_{\perp\beta}}{k^2}\frac{\bar{j}^2(k)}{|\Lambda(\mathbf{k_{\perp}\cdot u},k)|^2}\frac{dkd\phi}{k(2\pi)^2},
\end{align}
where in the first argument $\Lambda$ - functions there is a scalar product $\mathbf{k_{\perp}\cdot u}\equiv  ku\cos(\phi)\sin(\theta)$. Here $\theta$ is the angle between the directions of the particle velocity $\mathbf{v}$ and the drift velocity of the beam $\mathbf{u}$.
\par Since integration over the variable $\phi$ for an arbitrary function $|\Lambda(k,\theta,\phi)|^2$ is not possible, we approximate the corresponding value by its mean value:
\begin{align}
    \label{Another_aver_Diffus_tensor}    \hat{\mu}_{\alpha\beta}\approx\frac{1}{3}\Big(\frac{4\pi q}{mc^2} \Big)^2\sqrt\frac{\mathcal{E}}{2m}\int \frac{k_{\perp\alpha}k_{\perp\beta}}{k^2}\Big\langle\frac{\bar{j}^2(k)}{|\Lambda(k,\theta,\phi)|^2}\Big\rangle_{\phi}\frac{dkd\phi}{k(2\pi)^2}.
\end{align}
This substitution allows us to eliminate the complex dependence of the integrand on the axial angle $\phi$ and obtain simpler formulas for the diffusion coefficients, expressed as an integral only of the given spectral distribution function of currents. In fact, for the case we are interested in further, the inequality $\mathrm{Re}\ \varepsilon_{t}\ll$ $\mathrm{Im}\ \varepsilon_{t}$ holds and, therefore, the value $1/|\Lambda|^2$ can only vary between zero and one, so that the above averaging over the azimuthal angle can lead to a difference between the exact value of expression (\ref{Diffus_tensor_conv2}) and the approximate value (\ref{Another_aver_Diffus_tensor}) by only a numerical factor of the order of unity.
\par Finally, the tensor of coefficients of the FP equation will be written in the following form:
\begin{align}               
   \label{Diff_ten_struct_2}
   \hat{\mu}_{\alpha\beta}=\mu(\mathcal{E},\theta)\Big(\delta_{\alpha\beta}-\frac{{v}_{\alpha}v_{\beta}}{v^2}\Big),
\end{align}
\begin{align}               
   \label{Diff_coeff_2}
   \mu(\mathcal{E},\theta)\approx \frac{1}{6}\sqrt\frac{\mathcal{E}}{2m}\Big(\frac{4\pi q}{mc^2} \Big)^2\int_{0}^{+\infty} \Big\langle\frac{\bar{j}^2(k)}{|\Lambda(k,\theta,\phi)|^2}\Big\rangle_{\phi}\frac{dk}{k(2\pi)}.
\end{align}
\par A comparison of the expressions obtained (\ref{Diffus_tensor_struct}) and (\ref{Diff_ten_struct_2}) shows that for a time-independent magnetic field, the longitudinal coefficient $\mu_{||}$ is zero, which corresponds to the absence of particle diffusion in energy while preserving their diffusion in angles. Thus, as expected, particle scattering in a stationary magnetic field is completely elastic.
\par Further simplification of expression (\ref{Diff_coeff_2}) can only be achieved by assuming a specific form of the undisturbed distribution function that determines the dielectric permeability of the beam plasma $\varepsilon_{t}$. The presence of other non-collisional plasma, which is generally stationary relative to magnetostatic turbulence and supports it with its \say{external} currents, does not affect the result of screening the beam particles by their polarization currents. For the case of Maxwellian distribution of beam particles by velocity, the $\Lambda$-function in formula (\ref{Diff_coeff_2}) has the form \cite{akhiezerPlasmaElectrodynamics11975}:
\begin{align}
    \label{Dispersion_eq}
    \Lambda(\mathbf{k_{\perp}\cdot u},k)=1-\frac{(\mathbf{k_{\perp}\cdot u})^2}{k^2c^2}\Big\{1+i\sqrt{\frac{\pi}{2}}\frac{1}{k^2r_{D}^2}\frac{kv_T}{\mathbf{k_{\perp}\cdot u}}W\Big(\frac{\mathbf{k_{\perp}\cdot u}}{\sqrt{2}kv_T}\Big)\Big\}.
\end{align}
Here, $r_{D}$ is the Debye radius determined by the plasma beam, $v_{T}=\sqrt{T/m}$ is the average thermal velocity of particles, and $W(x)$ is the Faddeeva function (or Kramp function) \cite{landauCourseTheoreticalPhysics101981}. 
\par When the drift velocity of the beam is small compared to the thermal velocity of the particles, $u\ll v_{T}$ (Fig. \ref{fig:system}b), formula (\ref{Dispersion_eq}) reduces to the following asymptotic expression:
\begin{align}
    \label{Dispersion_eq_abs_lowlim}
    |\Lambda( k,\theta,\phi)|^2\approx 1+{\frac{\pi}{2}}\frac{1}{k^4r_{D}^4}\frac{v_T^2}{c^2}\Big(\frac{\mathbf{k_{\perp}\cdot u}}{k c}\Big)^2.
\end{align}
Using the approximate ratio (\ref{Dispersion_eq_abs_lowlim}) and averaging appropriately over the azimuthal angle $\phi$, we obtain an expression for the diffusion coefficient (\ref{Diff_coeff_2}) in the form
\begin{align}
    \label{non_aver_dif_coeff}
    \mu(\mathcal{E},\theta)\approx\frac{1}{6}\sqrt\frac{\mathcal{E}}{2m}\Big(\frac{4\pi q}{mc^2} \Big)^2\int_{0}^{+\infty} \frac{\bar{j}^2(k)}{\sqrt{1+\frac{\varkappa_{s}^4}{k^4}\sin^2(\theta)}}\frac{dk}{k(2\pi)},
\end{align}
where the characteristic wave number of the screening, determined by the plasma frequency of the beam, is introduced,
\begin{align}
    \varkappa_{s}=\Big(\frac{\pi}{2}\Big)^{1/4}\frac{\omega_{p}}{c}\sqrt{\frac{u}{v_T}}.
\end{align}
\par The screening length $l_{s}=1/\varkappa_{s}$ determined in this way depends on the average velocity of the directed motion of particles $u$ and decreases with its increase (this length also arises in the problem of shielding a magnetic dipole moving in plasma \cite{gubchenko1988generation}). The screening disappears, i.e. $\varkappa_{s}=0$, in the absence of relative motion between the beam and the sources of the turbulent field. This mode of interaction between the particle flow and magnetostatic fluctuations actually corresponds to the case of electromagnetic field penetration into plasma with an anomalous skin effect. \cite{landauCourseTheoreticalPhysics1981}. 
\par To simplify further consideration of the skin effect and obtain estimates in order of magnitude, we will also introduce the angle-averaged value of the diffusion coefficient $\theta$
\begin{align}
    \bar\mu(\mathcal{E})=\frac{1}{\pi}\int_{0}^{\pi}\mu(\mathcal{E},\theta)d\theta.
\end{align}
Performing this integration over the polar angle and moving from the spectral distribution of current density fluctuations $\bar{j}_{k}^2$ to the spectral distribution of magnetic fields $\bar{B}_{k}^2$, we obtain
\begin{align}
\label{average_din_coefficient}
    \bar\mu(\mathcal{E})=\frac{1}{6}\sqrt\frac{\mathcal{E}}{2m}\int_{0}^{+\infty} \frac{q^2\bar{B}^2_{k}}{m^2c^2}\frac{k^2}{\sqrt{k^4+\varkappa_{s}^4}}\frac{2}{\pi}K\Big(\frac{\varkappa_{s}^4}{k^4+\varkappa_{s}^4}\Big)\frac{kdk}{2\pi}.
\end{align}
Here, $K(x)$ is a complete elliptic integral of the first kind. From the form of expression (\ref{average_din_coefficient}), it follows that taking into account the finite length of the screening, $\varkappa_{s}\ne0$, leads to the appearance of an additional integrand factor, less than unity, which reduces the average value of the diffusion coefficient compared to its value for particle scattering in a magnetic field in a vacuum. 
\par If the spectrum of magnetic fluctuations is a strongly localized function with a sharp maximum near some value of the wave number $k_{opt}$, such that
\begin{align}
\label{spectr_B}
    \bar{B}^2_{k}\approx2\pi \bar B_{T}^2\tilde\delta(k-k_{opt}),
\end{align}
where $\tilde\delta(k-k_{opt})$ is the so-called physical delta function, which has a small but finite width, then the integral expression (\ref{average_din_coefficient}) can be approximately reduced to a simple formula
\begin{align}
\label{Mu_formula}
    \bar\mu(\mathcal{E})\approx \frac{1}{6}\bar\omega_{B}^2\sqrt\frac{\mathcal{E}}{2m} \frac{\lambda_{cor}}{\sqrt{1+\xi^4}}\frac{2}{\pi}K\Big(\frac{\xi^4}{1+\xi^4}\Big).
\end{align}
Here, the correlation length of magnetic turbulence $\lambda_{cor}=1/k_{opt}$, the dimensionless screening parameter $\xi=\lambda_{cor}\varkappa_{s}$, and the value $\bar{\omega}_ {B}=q\bar{B}_{T}/mc$ is the gyromagnetic frequency of the particles relative to the root mean square value of the turbulent magnetic field $\bar{B}_T$.
\par If the distribution of magnetostatic fluctuations has a large width in the wave vector space, then the effect of the shielding factor on different spectral components of the field will be different. According to the relation (\ref{non_aver_dif_coeff}), the effect of screening on the interaction of a particle beam with small-scale harmonics, which have small wavenumbers, $k\ll\varkappa_ {s}$, will be insignificant, and the scattering of particles will occur similarly to scattering in a magnetic field in a vacuum. In the opposite limit, $k\gg\varkappa_{s}$, the scattering will be significantly suppressed.
\par In the case of a large drift velocity, $u\gg v_{T}$, the effect of the screening factor on the diffusion coefficient remains qualitatively the same. However, it is not possible to simplify the analytical expression (\ref{Another_aver_Diffus_tensor}), since there is no any asymptotic expansion of the $\Lambda$-function in this limit. It can only be stated that for particle propagation angles $\theta$ such that $u\sin(\theta)\ll v_ {T}$, the effect of screening is always small, whereas when the reverse inequality is satisfied, the dispersion expression (\ref{Dispersion_eq}), accurate to terms of the order $u^2/c^2$, is equal to $1+1/(k^2\delta^2)$, where $\delta=c/\omega_{p}$ is the skin layer width. In the latter case, the shielding factor does not depend on the angle and is characterized only by the screening length, which is the skin length. Therefore, the corresponding mode, in contrast to the one described above, corresponds to the normal skin effect mode.   
\par Thus, based on the expressions (\ref{non_aver_dif_coeff}), (\ref{average_din_coefficient}) and  (\ref{Mu_formula}), we can conclude that the effect of screening on particle transport in the case of magnetostatic turbulence manifests itself in two ways: i) a decrease in the average value of the diffusion coefficient (see (\ref{Mu_formula})) and ii) the emergence of particle scattering anisotropy (see (\ref{non_aver_dif_coeff})) even for isotropic turbulence and in the absence of average values of electric and magnetic fields $\langle\mathbf{E}\rangle$ and  $\langle\mathbf{B}\rangle$, which is a consequence of the appearance in the system of a distinguished direction $\mathbf{e}_{||}\upuparrows\mathbf{u}$ associated with the motion of the beam as a whole.
%%%%%%%%%%%%%%%%%%%%%%%%%%%%%%%%%%%%%%%%%%%%%%%%%%%%%
\section{Anomalous plasma electrical conductivity. Numerical solution}
\label{sec4}
\par As discussed in previous sections, a turbulent magnetic field can result in anomalous plasma properties emerging. In particular, if scattering on magnetostatic perturbations supported by turbulent currents of another component (e.g. ions) is more significant than direct electron-ion collisions for one of the charged components (e.g. electrons), then anomalous electrical resistance arises. However, as discussed in Section \ref{sec3}, for a sufficiently dense particle flow, it is important to consider the screening effect, which modifies the scattering action of the field on them. The diffusion coefficients obtained from the FP equation above, which incorporate these changes, enable us to study the anomalous conductivity of plasma caused by magnetic turbulence.
\par In assumption of the given turbulence and negligible reverse action of the scattered particle beam (electrons) on external sources \footnote{This work does not address the issue of creating a turbulent field and maintaining a constant electric field $\mathbf{E}$
 (as well as the drift velocity $\mathbf{u}$).}, particle transport should obey the obtained kinetic equation of FP. In this paper, only the case of  small drift velocities, i.e., $u\ll v_{T}$,   homogeneous plasma, i.e., $\mathbf{v}\cdot \nabla \equiv 0$,  and the sufficiently narrow (see Section \ref{sec3}) spectral distribution of spatial fluctuations of the magnetic field is under consideration. Then equation (\ref{Fokker_Plank})  takes the form
\begin{align}
    \label{Kinetic_equation}
    \frac{\partial f_{0}}{\partial t}+\frac{q}{m}\mathbf{E}_d\cdot\frac{\partial f_{0}}{\partial \mathbf{v}}=\frac{1}{\sin(\theta)}\frac{\partial }{\partial \theta}\frac{D(\mathcal{E})\sin(\theta)}{\sqrt{1+\xi^4\sin^2(\theta)}}\frac{\partial f_{0}}{\partial \theta}.
\end{align}
Here, $\mathbf{E}_d$ is the average electric field that induces current and was mentioned in Section \ref{sec2}, and $D(\mathcal{E})=\bar\omega_ {B}^2\lambda_{cor}\sqrt{m/2\mathcal{E}}/12$ is the value of the diffusion coefficient in the velocity space in the absence of screening when particles move in a turbulent small-scale magnetostatic field. 
\par As can be seen from equation (\ref{Kinetic_equation}), the collision term on its right-hand side describes the diffusion of particles in angles, while keeping their energy unchanged, but leading to the isotropisation of the distribution function and a decrease in the magnitude of the directed flow velocity. The effective diffusion coefficient differs from the case of particle scattering when the influence of collective effects is neglected by the presence of a square root factor in the denominator, which, at a finite value of the screening parameter, $\xi\ne0$, leads to an angular dependence of the diffusion coefficient and, as a consequence, anisotropy of the transport properties of the plasma. In the approximation under consideration, the angular diffusion coefficient in configuration space can be compared to a uniaxial ellipsoid with the longest axis oriented along the direction of the particle drift velocity $\mathbf{e}_{||}\upuparrows\mathbf{u}$. The length of this axis, $D(\mathcal{E})$, gives the value of the diffusion coefficient along the flow direction, and the length of the smallest axis, $D(\mathcal{E})/\sqrt{1+\xi^4}$, determines its transverse value. It should be emphasised that screening actually only affects the reduction in the transverse diffusion coefficient, while its longitudinal value is the same as in a vacuum. 
\par The exact analytical solution of the kinetic equation (\ref{Kinetic_equation}) is quite a difficult task, and therefore only numerical solution is given. The presented results were obtained using standard grid methods for integrating the resulting differential kinetic equation in partial derivatives (see, for example, \cite{kahanerNumericalMethodsSoftware1989}). 
\begin{figure}    %%%%%%%%%%%%%%%%%% FIGURE 2
                          % includes the two top panels 
\centerline{\hspace*{0.015\textwidth}
         \includegraphics[width=0.515\textwidth,clip=]{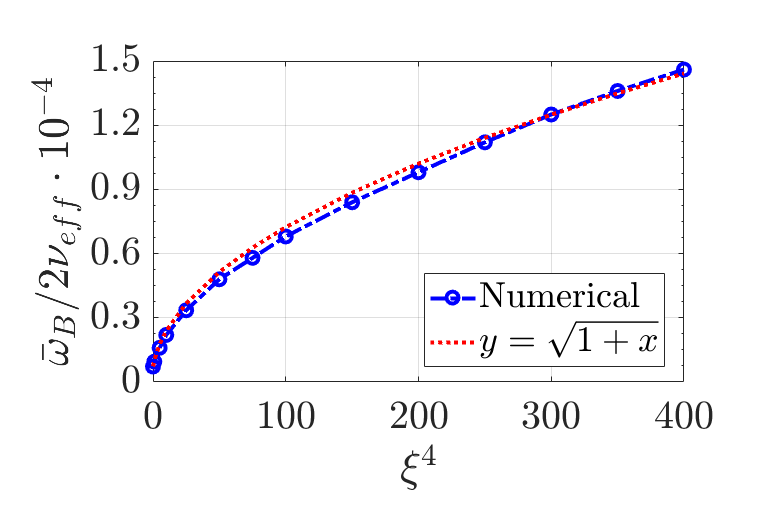}
         \hspace*{-0.03\textwidth}
         \includegraphics[width=0.515\textwidth,clip=]{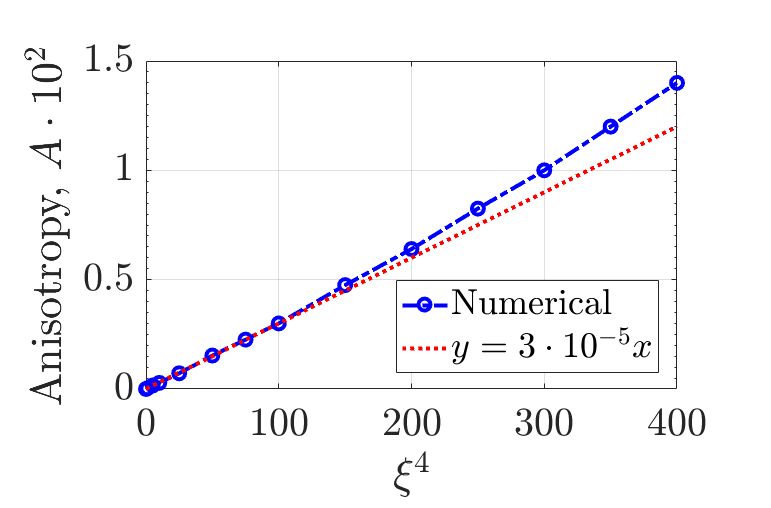}
        }
\vspace{-0.29\textwidth}   % Shift close to the panel top 
\centerline{\Large \bf     % Includes the labels (here needs the color 
                          %   package, see beginning of this file)
\hspace{0.13 \textwidth}  \color{black}{(a)}
\hspace{0.43 \textwidth}  \color{black}{(b)}
   \hfill}
\vspace{0.26\textwidth}    % Shift back to the panel          
\caption{Dependence of (a) dimensionless mobility $b$ and (b) anisotropy of the particle velocity distribution function $A$ on the screening parameter $\xi$. Numerical calculation - blue dashed-dotted curve, analytical approximation - red dotted curve.}
\label{fig:Anisotropy}
\end{figure}
%%%%%%%%%%%%%%%%%%%%%%%%%%%%%%%%%%%%%%%%%%%%%%%
\par Numerical solution to the FP equation is founded for various values of the screening parameter $\xi$ at a fixed level of turbulent fluctuations, i.e. the value $\bar{\omega}_{B}$, and a given external uniform electric field $\mathbf{E}_d\ne0$, which is not too large in magnitude: $E_d\ll E_{Dr}$, where $E_{Dr}\sim m\bar{\omega}^2_{B}\lambda_{cor}/q$ is the field value that determines the escape threshold of particles in turbulent plasma, similar to the Dreicer field \cite {dreicerElectronIonRunaway1959,dreicerElectronIonRunaway1960} for Coulomb collisions. The initial distribution of particles in the form of a Maxwellian function with a mean displacement ${u}_{init} < v_{T}$ are chosen. Under these conditions, there is a stationary solution for the distribution function $f_0$ with an average velocity $u$, corresponding to the flow of a constant electric current in plasma, the value of which is determined by the magnitude of the electric field and anomalous collisions, i.e. the scattering of particles on turbulence. As a result, finding the ratio of the dimensionless average particle velocity $u/(\sqrt{2}v_{T})$ to the value, also dimensionless, of the applied electric field $(E_d/\bar{B}_{T})\sqrt{2mc^2/T} $, one can obtain the dependence of the mobility of particles in plasma $b(\xi)=\bar{\omega}_{B}/2\nu_{eff}$ on the value of the screening parameter (here $\nu_{eff}=1/\tau_{c}$ is the effective collision frequency in turbulent plasma mentioned above, see Fig. \ref{fig:Anisotropy}a). 
\par As expected, strengthening the screening factor leads to an increase in the mean free path of particles $l_{c}$. According to Fig. \ref{fig:Anisotropy}a, the value of effective mobility is fairly well described by the formula $b=b_{0}\sqrt{1+\xi^4}$ (dashed curve in Fig. \ref{fig:Anisotropy}a), where the value $b_0$ determines the mobility of scattered in a turbulent magnetic field particles in the absence of screening. Nevertheless, the stationary particle velocity distribution function $f_0(\mathbf{v})$, which is the solution to the FP equation, differs from the Maxwellian distribution, as expressed by the appearance of a non-zero anisotropy parameter $A=1-T_{||}/T_{\perp}$ , where $T_{||}$ and $T_{\perp}$ are the longitudinal and transverse temperatures of the steady flow. The dependence of the anisotropy value on the parameter $\xi$ is shown in Fig. \ref{fig:Anisotropy}b.
\par Thus, even for isotropic turbulence arising in the presence of a distinct direction of mean particle motion, the anisotropy of the diffusion coefficient leads to the appearance of anisotropy in the velocity distribution function, which, when the screening parameter is not too large, increases inversely proportional to the fourth power of the screening length $l_s$ (dashed line in Fig. \ref{fig:Anisotropy}b). Note that in the absence of additional isotropization factors, i.e., when the anisotropy of the particle distribution is preserved for sufficiently long times, it can become a source of kinetic plasma instabilities, such as Weibel \cite{weibelSpontaneouslyGrowingTransverse1959,hasegawaPlasmaInstabilitiesNonlinear1975, kocharovskyAnalyticalTheorySelfconsistent2016,emelyanovWeibelInstabilityPresence2024}, which in turn leads to the generation of a small-scale turbulent magnetic field that provides momentum exchange between electrons and ions. In other words, in collisionless plasma in the absence of strong quasi-electrostatic pulsations, it is possible to realise a self-sustaining mode of electric current flow, for which the conductivity is determined by magnetic field fluctuations. Such a situation can occur under various conditions in laboratory or space plasma, for example, in the central part of current layers \cite{Vicentin_2025,nechaev2025turbulent} during magnetic reconnection.   
\par In the case of sufficiently large values of the external electric field, $E_d\sim E_{Dr}$, as with Coulomb collisions, there is no steady-state solution and particles escape, continuously accelerating and gaining energy. Preliminary numerical calculations show that this causes a strong deformation of the distribution function and an increase in anisotropy, $T_{\perp}>T_{||}$. A correct analysis of this regime requires the calculation of the self-consistent change in the distribution function and the screening parameter, which depends on the average beam velocity. In addition, if the directional velocity is comparable in magnitude to the thermal velocity, $u\gtrsim v_{T}$, it is necessary to go beyond the simple approximation considered above and take into account the screening factor determined by the ratio (\ref{Dispersion_eq}) more accurately. Moreover, at sufficiently high relative velocities of different plasma components, instabilities may develop, leading to the generation of other types of turbulence not considered in this work \cite{cytovicIntroductionTheoryPlasma1972,hasegawaPlasmaInstabilitiesNonlinear1975, akhiezerPlasmaElectrodynamics11975}. 
%%%%%%%%%%%%%%%%%%%%%%%%%%%%%%%%%%%%%%%%%%%%%%%%%%%%%
\begin{figure}    %%%%%%%%%%%%%%%%%% FIGURE 3
                          % includes the two top panels 
\centerline{\hspace*{0.015\textwidth}
         \includegraphics[width=0.515\textwidth,clip=]{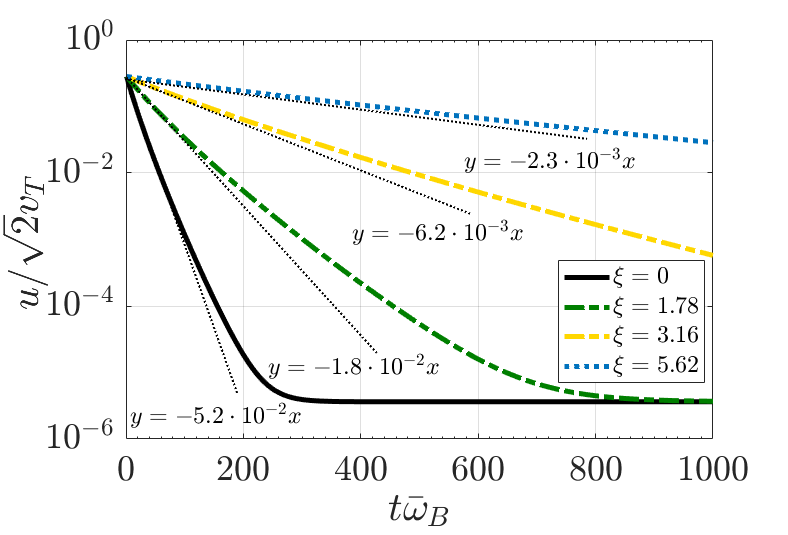}
         \hspace*{-0.03\textwidth}
         \includegraphics[width=0.515\textwidth,clip=]{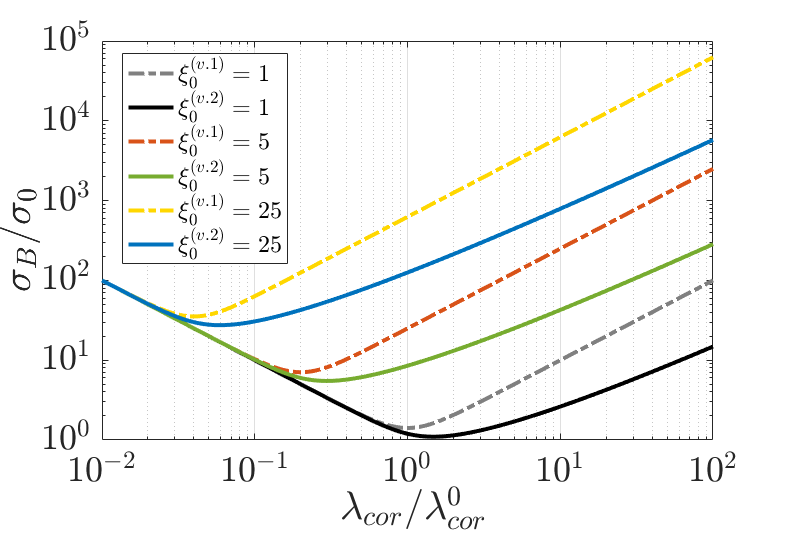}
        }
\vspace{-0.31\textwidth}   % Shift close to the panel top 
\centerline{\Large \bf     % Includes the labels (here needs the color 
                          %   package, see beginning of this file)
\hspace{0.34 \textwidth}  \color{black}{(a)}
\hspace{0.42 \textwidth}  \color{black}{(b)}
   \hfill}
\vspace{0.31\textwidth}    % Shift back to the panel          
\caption{a) Time dependence of the normalized average particle velocity $u$ during free relaxation of its initial value for different values of the screening parameter $\xi$. b) Dependence of the normalized effective conductivity of turbulent plasma $\sigma_{B}$ on the characteristic scale of magnetic perturbations $\lambda_{cor}$. The parameter $\lambda_{cor}^{0}$ is equal to the screening length $l_s$ at $\xi=1$.}
\label{fig:Diffuse_Drift}
\end{figure}
%%%%%%%%%%%%%%%%%%%%%%%%%%%%%%%%%%%%%%%%%%%%%%%
\par It should be noted that for real plasma, the applicability of the obtained results may be limited by the finite free path length of particles, which is not related to the magnetic fluctuations under consideration, and by the dimensions of the system or the scale of plasma inhomogeneity.   
\par  Turning to the problem of free relaxation of a particle beam with a given initial drift velocity $u$, when there is no external average electric field $\mathbf{E}_d$, we can also estimate the free path time of particles in turbulent plasma based on the estimate of the characteristic decay time of the initial current value \footnote{It should be noted that the application of the analytical formulas obtained in the previous section to describe the relaxation of the initial particle beam velocity assumes that the characteristic time of change of the averaged distribution function $f_0$, determined by the effective collision frequency $\ nu_{eff}$, is much greater than the characteristic time of magnetic field fluctuations in the beam reference frame, i.e. $\nu_{eff}\ll k_{opt}u$. This condition imposes a restriction on the intensity of magnetic turbulence: $\lambda_{cor}^2/\bar{r}^2_{H}\ll (u/v_{T})\sqrt{1+\xi^4}$.}. Fig. \ref{fig:Diffuse_Drift}a shows the numerically determined time dependencies of the normalized drift velocity $u/(\sqrt{2}v_T)$ for different values of the screening parameter $\xi$. The initial sections of the curves in this figure correspond to the stage of exponential decay of currents, and the characteristic decay rate determines the average frequency of particle collisions $\nu_{eff}$, and thus their mobility. However, in the case under consideration, the dependence of the effective mobility of particles $b$ on the screening parameter is not described by the simple formula $b=b_{0}\sqrt{1+\xi^4}$ mentioned above. In reality, this dependence corresponds to that given by expression (\ref{Mu_formula}), where a factor with an elliptic integral is additionally taken into account:
\begin{align}
\label{mobility}
    b=b_{0}\frac{\pi\sqrt{1+\xi^4}}{2 K(\xi^4/(1+\xi^4))}.  
\end{align}
\par As can be seen from the above, determining the free path length of particles in a plasma with a turbulent magnetic field is not straightforward. However, the change in particle mobility due to the screening effect is qualitatively the same in all cases of weak turbulence. 
\par Knowing the corresponding value of effective mobility (or, equivalently, collision frequency $\nu_{eff}$), one can estimate the anomalous plasma conductivity $\sigma_B$ caused by particle scattering in a turbulent magnetic field. According to the classical Drude formula, for the first and second methods of estimating the screening effect considered above (with and without an average electric field $\mathbf{E}_d$), we have, respectively: 
\begin{align}
\label{conductivity}
    \sigma_B=\frac{e^2n}{m\nu_{eff}}=\frac{2e^2n}{m\bar{\omega}_{B}}b\sim\frac{e^2n}{m\bar{\omega}_{B}^2}\frac{v_{T}}{\lambda_{cor}}\sqrt{1+\lambda_{cor}^4\varkappa_{s}^4} \text{ или }\frac{e^2n}{m\bar{\omega}_{B}^2}\frac{v_{T}}{\lambda_{cor}}\frac{\pi\sqrt{1+\lambda_{cor}^4\varkappa_{s}^4}}{2K(\lambda_{cor}^4\varkappa_{s}^4/(1+\lambda_{cor}^4\varkappa_{s}^4))}.
\end{align}
\par Fig. \ref{fig:Diffuse_Drift}b shows the dependence of the normalized value of the effective plasma conductivity $\sigma_B/\sigma_0$ (here $\sigma_0$ is a dimensional factor in formula (\ref{conductivity}) that does not depend on $\lambda_{cor}$) on the characteristic correlation length of magnetostatic perturbations $\lambda_{cor}$ for the first and second cases ($v.1$ and $v.2$, respectively). The curves obtained indicate that different turbulence scales make different contributions to conductivity. For magnetic turbulence with a broad spectrum, which can arise, for example, as a result of a turbulent cascade during non-linear energy transfer between different spectral components, the largest contribution to particle scattering is made by inhomogeneities with a scale of the order of the screening length, $\lambda_{cor}\sim l_s$. Thus, this value determines the characteristic scale at which magnetic field fluctuations interact most effectively with electrons (or ions, if turbulent currents are supported by electrons), i.e., the scale of turbulence dissipation. However, a detailed study of this phenomenon requires a more consistent description of magnetic turbulence itself, including consideration of the coordinated change in the field and the distribution of particles by energy.  
\par The numerical and analytical results presented above are intended only to illustrate the general patterns of the non-linear screening effect of scattering. They do not take into account a number of factors that may affect the specific numerical values of the calculated diffusion coefficients and anomalous conductivity values, in particular, they do not address the issue of one or another type of magnetic turbulence spectrum. More accurate calculations, including those using the particle-in-cell method \cite{borodachevDynamicsSelfConsistentMagnetic2017}, are the subject of a separate study. 
%%%%%%%%%%%%%%%%%%% SECTION_5 %%%%%%%%%%%%% %%%%%%%%%%%%%%%%%%%%%%%%%%%%%%%%%%%%%%%%%%%
\section{Conclusion}
\label{sec5}
\par This paper examines how the electromagnetic screening effect influences anomalous particle transport in dense, non-relativistic, collisionless plasma with a microturbulent magnetic field that has a zero mean vector and satisfies the weak turbulence condition. Using a quasi-linear kinetic description, general expressions for the diffusion coefficients of particles in velocity space that account for this effect are obtained. The case of magnetostatic fluctuations is considered separately, and simple analytical formulas are derived for the relevant coefficients in the FP equation, as well as for the anomalous plasma conductivity caused by particle scattering in a turbulent magnetic field. These formulas explicitly account for the screening effect, demonstrating that it not only decreases the average values of the kinetic coefficients and increases the anomalous conductivity, but also leads to the emergence of anisotropy in the transport properties of the plasma. This is due to the appearance of a distinguished direction in the system when the average velocity of a particle beam is present.  
\par The paper also presents the results of a numerical solution to the FP equation, which describes the angular diffusion of particles propagating in a plasma with a small-scale turbulent magnetic field. It demonstrates that, in the presence of an external constant electric field, which establishes an average electric current, the stationary distribution function becomes anisotropic one, and the anisotropy parameter increases in inverse proportion to the fourth power of the screening length. 
\par  The results obtained were compared with those previously known for low-density plasma flows, for which the screening effect of magnetic turbulence scattering is negligible. However, important questions remain unanswered, such as providing a self-consistent description of the spectrum of turbulent pulsations and particle dynamics. This would require taking into account changes in the screening length when the distribution function changes and describing the reverse effect of scattered particles on external sources that support magnetic field fluctuations. This paper does not consider the case where the turbulent field is non-stationary and particle diffusion by energy is possible in detail. In this case, screening may be significant, potentially leading to anisotropy in the heating and acceleration of particles by turbulent pulsations.  
\par The influence of the screening effect on anomalous particle transport in magnetically active plasma with a non-zero mean magnetic field vector is of great interest. Under these conditions, the resonant interaction of turbulent magnetic pulsations with various normal oscillations and waves in the plasma may also be important and should be taken into account. The case of strong turbulence is of particular interest, as there is a significant proportion of magnetized particles in this scenario \cite{islikerTransportParticlesStrongly2025}. Further research in this area is desirable.
   
%%%%%%%%%%%%%%%%%%% BIBLIOGRAPHY %%%%%%%%%% %%%%%%%%%%%%%%%%%%%%%%%%%%%%%%%%%%%%%%%%%%%

\textbf{Acknowledgements.}   This work was supported by a Russian Science Foundation grant No 22-12-00308-P.
\bibliography{biblio}
\end{document}